\RequirePackage{fix-cm}
\documentclass[smallextended]{svjour3}       % onecolumn (second format)
\smartqed  % flush right qed marks, e.g. at end of proof
\usepackage{graphicx}
\usepackage{subfig}
\usepackage{natbib}
\usepackage{url}

\journalname{Experimental Astronomy}
\begin{document}

\title{Stellar masks and bisector's shape for M-type stars observed in the
GAPS Programme with HARPS-N at TNG
\thanks{Based on observations made with the Italian Telescopio Nazionale
Galileo (TNG) operated on the island of La Palma by the Fundacion
Galileo Galilei of the INAF at the Spanish Observatorio Roque de los
Muchachos of the IAC in the frame of the program Global Architecture
of the Planetary Systems (GAPS).}
}

\titlerunning{Stellar masks for M-type stars}        % if too long for running head

\author{Monica Rainer         \and
        Francesco Borsa \and
        Laura Affer 
}


\institute{M. Rainer \at
              INAF-OAA, Largo E. Fermi 5, 50125 Florence, Italy \\
              Tel.: +39-055-2752205\\
              \email{monica.rainer@inaf.it}           %  \\
           \and
           F. Borsa \at
              INAF-OAB
           \and
           L. Affer \at 
              INAF-OAPa
}

\date{Received: date / Accepted: date}

\maketitle

\begin{abstract}
   The HARPS/HARPS-N Data Reduction Software (DRS) relies on the cross-correlation between the
   observed spectra and a suitable stellar mask to compute a cross-correlation function
   (CCF) to be used both for the radial velocity (RV) computation and as an indicator of
   stellar lines asymmetry, induced for example by the stellar activity.
   Unfortunately the M2 mask currently used by the HARPS/HARPS-N DRS for M-type stars
   results in heavily distorted CCFs.

   We created several new stellar masks
   in order to decrease the errors in the RVs and to improve the
   reliability of the activity indicators as the bisector's span.

   We obtained very good results with a stellar mask created from the theoretical
   line list provided by the VALD3 database for an early M-type star
   (T$_{\mathrm{eff}}$=3500~K and $\log{g}=4.5$).
   The CCF's shape and relative activity indicators improved and the RV time-series
   allowed us to recover known exoplanets with periods and amplitudes compatible with
   the results obtained with HARPS-TERRA.
\end{abstract}

\section{Introduction}
\label{intro}
   Even after the advent of ESPRESSO \citep{2010SPIE.7735E..0FP},
   the twin high-resolution echelle spectrographs HARPS \citep{2003Msngr.114...20M}
   and HARPS-N \citep{2012SPIE.8446E..1VC}, installed at the 3.6m telescope at the
   ESO-LaSilla Observatory and at the Telescopio Nazionale Galileo (TNG) at the 
   Roque de los Muchachos Observatory respectively, remain two of the most important
   instruments for radial velocity (RV) measurements.
   They are mainly used for exoplanets search and characterization, an
   ever evolving field that has recently focused on the M-type stars \citep[e.g.][]{2013A&A...549A.109B,
   2016A&A...593A.117A}.
   These cool stars are very interesting targets for exoplanetary studies,
   due to their long lives, low masses and close-in habitable zone \citep[e.g.][]{2013A&A...553A...8D,
   2013A&A...549A.109B,2014MNRAS.441.1545T,2015ApJ...807...45D}
   but they present the strong drawback of showing usually very high activity levels,
   which results in large quasi-periodic variations in the RV time-series
   \citep{2014Sci...345..440R,2016ApJ...830...74A}.
   As such, it is very important to have reliable stellar activity indicators when
   looking for planets orbiting M-type stars, in order to discriminate between
   the planetary Keplerian signals and the activity signals.

   Both HARPS and HARPS-N data are processed by a Data Reduction Software (DRS) optimized
   for exoplanet search \citep{1996A&AS..119..373B,2002A&A...388..632P},
   that computes the RVs by cross-correlating the spectra
   with a suitable mask chosen from a template library. The mask consists in a list of wavelength
   ranges (identifying the spectral lines) and weights used to define the contribution
   of each single spectral line to the cross-correlation. 
   The resulting cross-correlation
   function (CCF from here on) is used not only to compute the RV,
   but also to build some activity indicators, as the bisector's span (an indicator of the
   asymmetry of the CCF, denoting the curvature of the line that bisects the CCF), the CCF
   contrast (i.e. the depth of the CCF at its central point) and full-width at half-maximum
   (FWHM) of the CCF.

\begin{figure*}
   \subfloat{\includegraphics[width=0.5\textwidth]{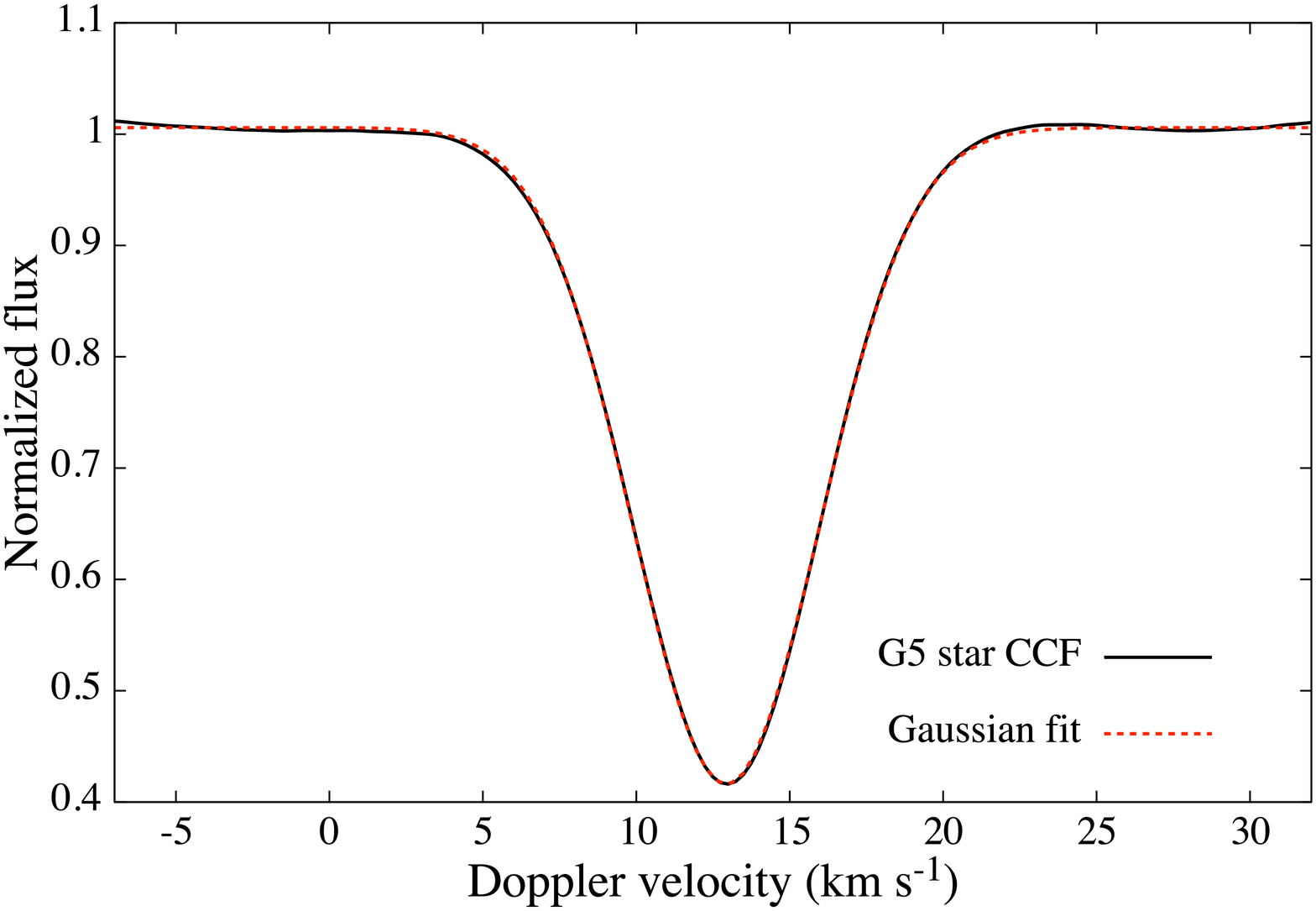}}
   \subfloat{\includegraphics[width=0.5\textwidth]{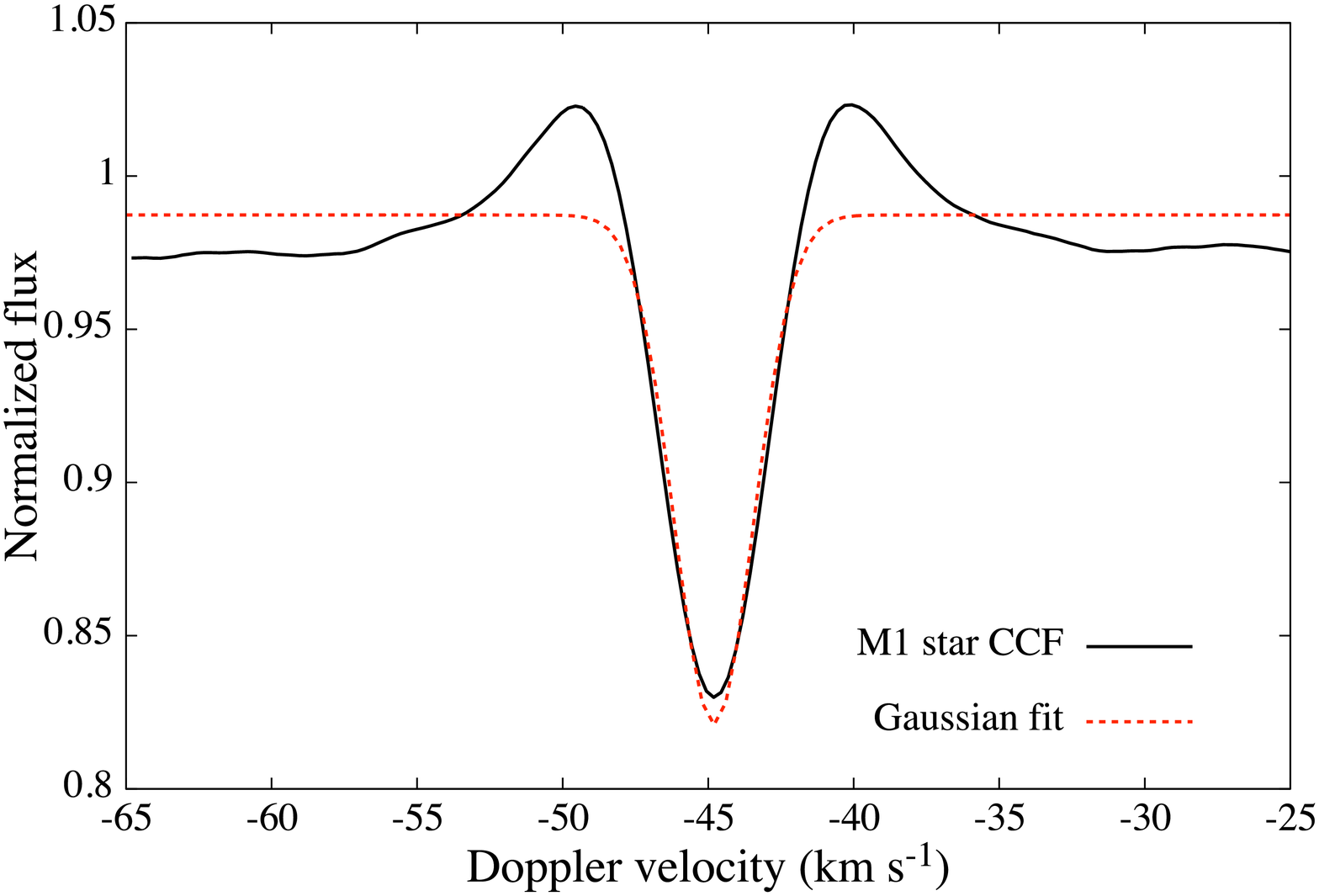}}
   \caption{\textit{Left:} CCF created by the online DRS for a G5 star using the G2 mask and
   relative Gaussian fit.
   \textit{Right:} CCF created by the online DRS for a M1 star using the M2 mask and relative Gaussian fit.}
   \label{gaussian_fit}
   \end{figure*}

   Unfortunately the online pipeline running at the telescope uses a very limited mask library,
   consisting of only three masks, optimized for G2, K5 and M2-type stars.
   In the case of M-type stars using the M2 mask, i.e. the most suitable
   mask for those stars, gives less satisfactory results than those obtained when
   processing solar-like stars with the G2 mask, in terms of both the shape of the
   CCF and the resulting RVs, which are computed using a Gaussian fit
   (see Fig.~\ref{gaussian_fit}).
   The resulting CCFs deviate from a Gaussian profile and show
   notable bumps on the wings (see Sect.~\ref{mask01}), while the bisector's shape
   does not seem to map the stars' velocity fields (see Sect.~\ref{bis_shape}).
   Many studies avoid this issue by using the RVs derived by the HARPS-TERRA software
   \citep{2012ApJS..200...15A, 2017A&A...598A..26P}, that performs a 
   least-squares matching of each observed spectrum in a time-series to a high 
   signal-to-noise ratio template derived from the same dataset. Still, in order
   to recover some activity indicators, many scientists still use the DRS bisector's span
   \citep{2014MNRAS.443L..89A,2014ApJ...791..114W},
   or they compute their own bisector's span on the DRS CCF
   \citep{2015ApJ...805L..22R,2018Natur.553..477B}.

   The aim of this paper is to show how the use of different custom-made stellar masks
   for M-type stars gives very different results in terms of the shape of the CCFs,
   the RV precision and scatter, and the bisector's shape.
   We will illustrate how both the RV measurements and the use
   of the bisector's span as an activity indicator may be improved by using a
   better defined mask for early M-type stars.

\section{Spectroscopic data}
\label{sec:1}
   For our work we focused on the HARPS-N data, because we were able to re-reduce them
   many times with different ad-hoc masks using the YABI platform
   \citep{2012SourceCodeBiolMed...7...1} hosted at IA2 Data Center\footnote{\url{https://www.ia2.inaf.it}}
   in the mainframe of the italian GAPS
   (Global Architecture of Planetary Systems) program \citep{2013A&A...554A..28C,2016MmSAI..87..141P},
   but we assume the same results are valid for the HARPS data, given that both the
   instruments and the reduction pipelines are twins.
   We already used this method in the past in order to use personalized stellar mask
   for particular targets, as in the case of the F6V star $\tau$~Bootis \citep{2015A&A...578A..64B},
   where the standard G2 mask used by the DRS under-performed due to the mis-matched stellar type.

   In this work, we studied the effect of different masks on the CCF's shape, RV
   determination and bisector's shape of early M-type stars using several datasets
   of HARPS-N spectra.
   We used the time-series of the exoplanet host stars GJ~3998 \citep{2016A&A...593A.117A}
   and GJ~625 \citep{2017A&A...605A..92S} to
   create new stellar masks, and then we tested their results in term of RVs and bisector's
   span against the results obtained using the standard M2 and K5 masks. We also compared
   the RVs those obtained using the HARPS-TERRA software on both these targets
   and two other additional exoplanet host stars, GJ~686 \citep{2019A&A...622A.193A}
   and GJ~3942 \citep{2017A&A...608A..63P}.

   The data used in this work are:
   \begin{itemize}
   \item 195 spectra of GJ~3998 (M1V, V=10.83 mag) observed from 26 May 2013 to 28 August 2018;
   \item 164 spectra of GJ~625 (M1.5V, V=10.17 mag) observed from 25 May 2013 to 12 September 2017;
   \item 64 spectra of GJ~686 (M1.5Ve, V=9.577 mag) observed from 11 February 2014 to 20 October 2017;
   \item 146 spectra of GJ~3942 (M0.5Ve, V=10.25 mag) observed from 25 May 2013 to 27 February 2017.
   \end{itemize}
 
   The stars were observed as part of the HADES (HArps-n red Dwarf Exoplanet Survey)
   program, a collaboration between the GAPS consortium, the Institut de Ciències
   de l’Espai de Catalunya (ICE), and the Instituto de Astrofísica de Canarias (IAC).

\section{Cross-correlation function and stellar masks: first approach} \label{mask01}
\label{sec:2}
   \begin{figure*}
   \resizebox{\hsize}{!}{\includegraphics[trim={0 0 0 2.5cm},clip]{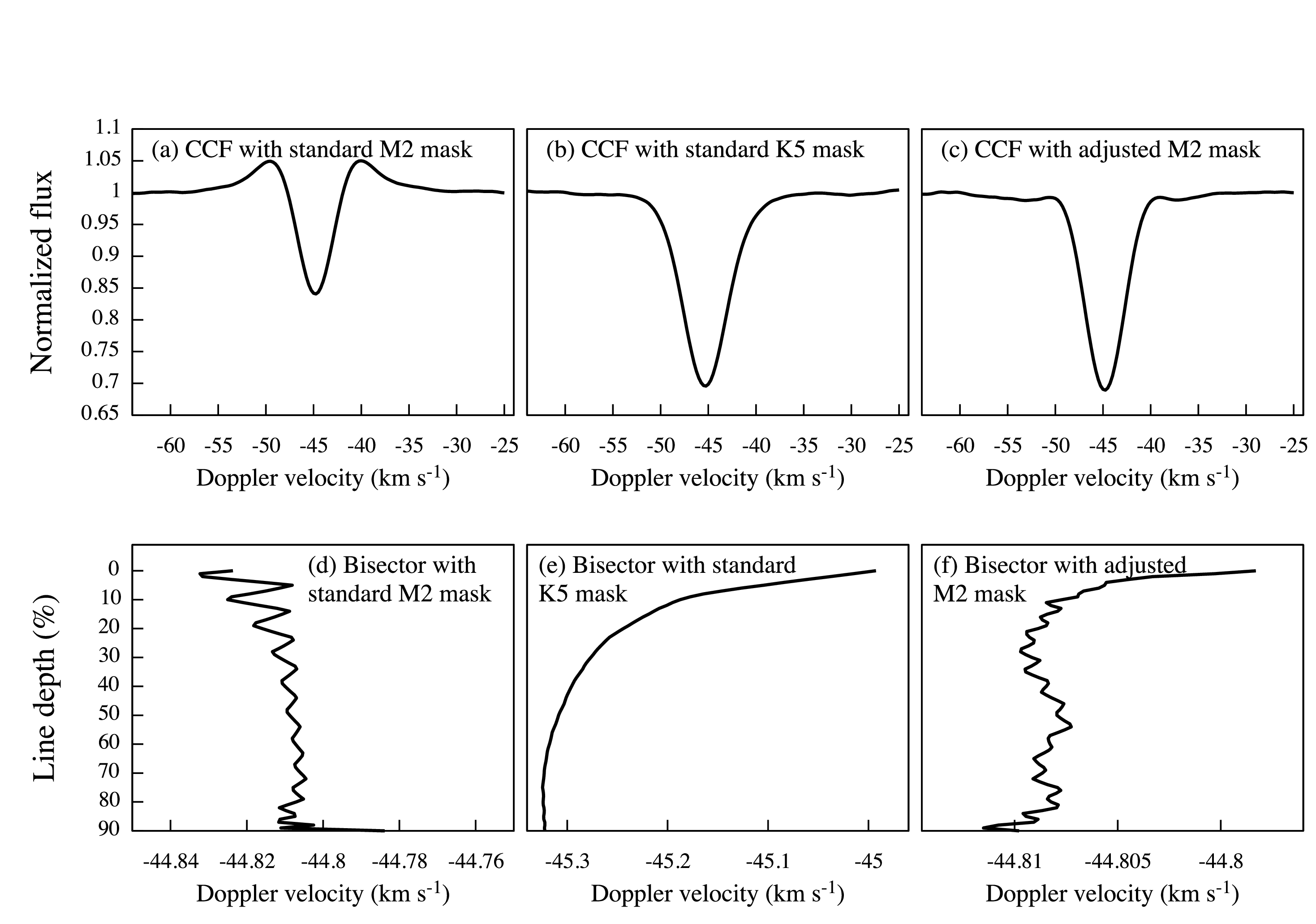}}
   \caption{Examples of CCFs (upper row) and bisectors (lower row) of GJ~3998 obtained with
   the M2 (left), K5 (middle) and adjusted M2 (right) mask respectively.}
   \label{ccfbis01}
   \end{figure*}

   The HARPS-N CCF for the early M-type GJ~3998 star obtained using the M2 mask shows
   an immediate drawback: the line profile does not resemble a Gaussian (as it is expected
   in the case of slow-rotating stars), but it presents
   two bumps on the wings (see Fig.~\ref{ccfbis01}$a$, upper left panel).
   These bumps are present in the CCFs of any M-type star that is reduced using the M2 mask,
   so they are not a feature of this particular object.
   This effect is particularly troubling considering that the HARPS-N pipeline uses a Gaussian
   fit of the CCF to compute the radial velocity of the star.
   The bumps disappear using the K5 mask (see Fig.~\ref{ccfbis01}$b$, upper middle panel),
   but the mean RV error doubles (see Tab.~\ref{vrad_bis_rms}).
   We note here that the rms in both the RV and the bisector's span increases using the K5 mask.
   \begin{figure*}
   \subfloat{\includegraphics[width=0.5\textwidth,trim={0 0 0 4.5cm},clip]{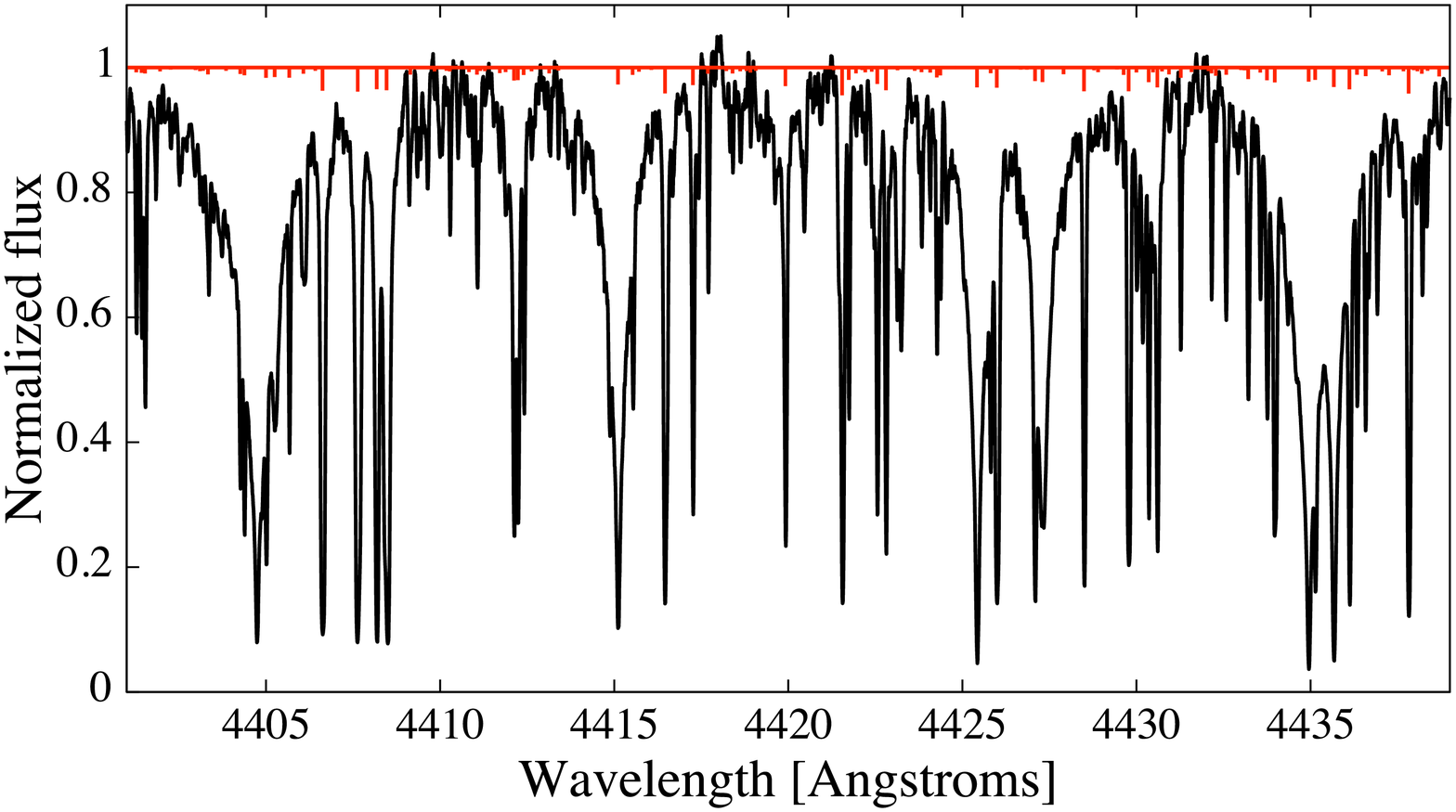}}
   \subfloat{\includegraphics[width=0.5\textwidth,trim={0 0 0 4.5cm},clip]{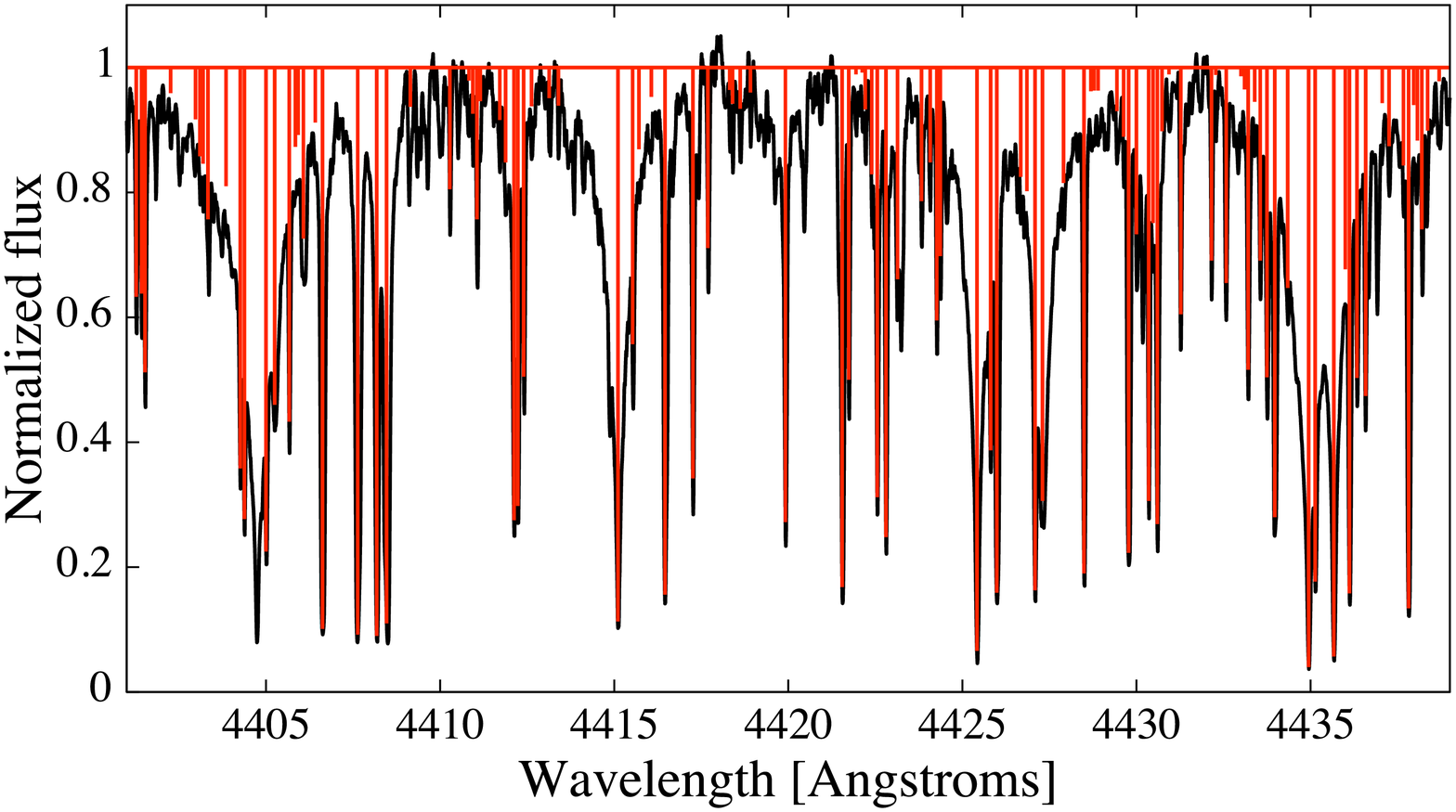}}
   \caption{Comparison between the original (left) and adjusted (right) M2 mask
    with an observed spectrum.}
   \label{masks}
   \end{figure*}

   As a first approach, we looked at the M2 mask by overplotting it over an observed
   early M-type star (see Fig.~\ref{masks}, left panel).
   We found a first obvious issue: all the lines in the mask have mostly the same 
   very shallow depths, which are used by the pipeline to weigh their contribution
   to the CCF. This means that all the 9196 lines of the mask contribute roughly in the same
   way to the CCF, regardless of their intensity.

   We decided to modify the M2 mask by adjusting the line depths to the values of our observed
   spectrum and by eliminating the most shallow lines (see Fig.~\ref{masks}, right panel), i.e.
   lines with depths less than 2.5\% of the continuum. This threshold value was chosen in order to
   get rid of lines that would only contribute to the noise of the CCF (particularly considering
   the difficulty in defining the continuum in M-type stars) while maintaining a large enough number
   of lines in the stellar mask to obtain a reasonable CCF.
   We ended up with a number of 5586 usable lines, much less than the original one, but comparable to those
   of the other DRS masks (e.g., the K5 mask consists of 4828 lines).

   The use of this new mask partially improved our qualitative results: the CCF's shape
   lost the bumps
   and more closely resembled a Gaussian (see Fig.~\ref{ccfbis01}$c$, upper right panel),
   but the RV mean error remained more or less the same as the one obtained
   with the M2 mask, as did the rms of both the RV and the bisector span
   (see Tab.~\ref{vrad_bis_rms}).

   We tested the new mask also on the M-type star GJ~625, obtaining the same qualitative results.

   \begin{table}
   \caption{Summary results of RVs errors and dispersion obtained with different masks}             % title of Table
   \label{vrad_bis_rms}      % is used to refer this table in the text
   \centering                          % used for centering table
   \begin{tabular}{c c c c}        % centered columns (4 columns)
   \hline\hline                 % inserts double horizontal lines
   Mask & RV mean error & RV rms & Bis. span rms \\    % table heading
    & (m s$^{-1}$) & (m s$^{-1}$) & (m s$^{-1}$) \\    % table heading 
   \hline                        % inserts single horizontal line
    & \multicolumn{3}{c}{GJ~3998} \\
   \hline                        % inserts single horizontal line
      M2 & 1.88 & 4.89 & 3.75 \\      % inserting body of the table
      K5 & 3.59 & 6.46 & 9.75 \\
      Adj. M2 & 2.00 & 4.78 & 4.51 \\
      MVALD & 1.98 & 4.81 & 4.46 \\
      Adjusted MVALD & 2.06 & 4.89 & 4.93 \\
      Deblended MVALD & 3.89 & 6.82 & 6.95 \\ 
   \hline                        % inserts single horizontal line
    & \multicolumn{3}{c}{GJ~625}\\
   \hline
      M2 & 1.13 & 2.78 & 2.29 \\      % inserting body of the table
      K5 & 3.13 & 6.82 & 13.82 \\
      Adj. M2 & 1.30 & 3.21 & 3.34 \\
      MVALD & 1.22 & 3.51 & 3.22 \\
      Adjusted MVALD & 1.29 & 3.67 & 3.60 \\
      Deblended MVALD & 2.31 & 4.65 & 3.69 \\ 
   \hline                                   %inserts single line
   \end{tabular}
   \end{table}

%-----------------------------------------------------------------

\section{Bisector's shape} \label{bis_shape}
\label{sec:3}
   Searching for exoplanets around M-type stars requires reliable activity indicators,
   due to the stars' high activity levels. Many M-type stars are faint objects, and as
   such the information that can be derived from the CCF are important: the CCF is
   obtained by combining a large number of spectral lines, allowing us to reach a high
   signal-to-noise ratio even for faint objects.

   The CCF's bisector is an useful indicator for the presence of different velocity fields
   in the star. Convection in solar-like stars results in a net convective blueshift,
   arising from the different contributions of the ascending granules and the descending
   intergranular lanes. The two components combine in a slightly asymmetric line profile,
   that translates to a C-shaped bisector in quiet stars \citep{2005oasp.book.....G}.
   The magnetic activity will change and modulate the bisector's shape
   \citep{1999ASPC..185..268D}, as phenomena such as faculae and plage suppress part of
   the convective blueshift. As such, the bisector's time-series may show the activity
   cycles' progression.

   The effects of the activity on the shape of the spectral lines of M-type stars are still
   not very well known, even if the RVs variations of active M dwarfs have been studied
   \citep{2018A&A...614A.122T}. In principle, a similar behaviour is expected in the bisector's shape
   and its variations.

   We note here that usually the bisector's shape is not used directly as an
   activity indicator:
   the bisector's span stands as its proxy and it is defined as the difference between
   the mean values of the bisector in the upper and lower part of the CCF, giving an
   immediate insight in the asymmetry of the stellar lines. Comparing the radial
   velocity time-series with the bisector's span time-series allows to discriminate
   between Keplerian and stellar variations.

   We compared the shapes of the bisectors of the CCFs obtained using the original
   M2 mask and the K5 mask (see Fig.~\ref{ccfbis01}$d$ and Fig.~\ref{ccfbis01}$e$,
   lower left and middle panel respectively): while the K5 bisector shows the
   expected C-shape, the M2 bisector seems to tilt in the opposite direction and
   it is afflicted by some non-trivial deformities, which can hardly be explained
   in some physical way.
   The adjusted M2 mask smoothes the bisector, but the shape is still quite different
   from the K5 bisector and some of the segmentation from the original M2 mask remains
   (see Fig.~\ref{ccfbis01}$f$, lower right panel).

   The bisectors obtained using either the original or the modified M2 mask seem to have
   little physical meaning and they should not be used for any kind of activity study.

   We decided to follow a different strategy and to build a new mask from scratch.

%-----------------------------------------------------------------

\section{Cross-correlation function and stellar masks: second approach} \label{mask02}
\label{sec:4}
   Instead of modifying an already existing stellar mask or building a stellar mask from an
   observed spectrum (which in the case of M-type stars would be a long and difficult
   project due to their numerous and heavily blended lines), we decided to use as
   a starting point a single theoretical line list. In order to do so, we queried the VALD3
   database\footnote{http://vald.astro.univie.ac.at/$\sim$vald3/php/vald.php}
   \citep{1995A&AS..112..525P} and downloaded the line list for a star with
   T$_{\mathrm{eff}}$=3500~K (the lowest available temperature in the VALD3 database)
   and $\log{g}=4.5$ in the wavelength region 3900-7000{\r{A}}.

   We transformed the line list in the right format required by the HARPS-N pipeline, then we
   cut the Balmer lines' regions and the regions where most of the telluric lines are found,
   because the former may influence the CCF's width and the latter may introduce spurious
   RV modulations \citep{2002A&A...388..632P}.
   Then we removed the most shallow lines, that mostly introduce
   noise, and we obtained finally the MVALD mask.
   It contains 17547 lines, almost double than the number of the
   lines of the original M2 mask (9196), spanning over the wavelength regions
   4400-5027~\r{A}, 5105-5303~\r{A}, 5337-5405~\r{A}, 5504-5574~\r{A},
   5580-5675~\r{A}, 5770-5858~\r{A}, 6018-6267~\r{A}, 6374-6432~\r{A},
   6618-6862~\r{A}. We then adjusted the depths of the lines using
   an observed spectrum of GJ~3998 to create an adjusted MVALD mask. 

\begin{figure*}
   \resizebox{\hsize}{!}{\includegraphics[trim={0 0 0 2.5cm},clip]{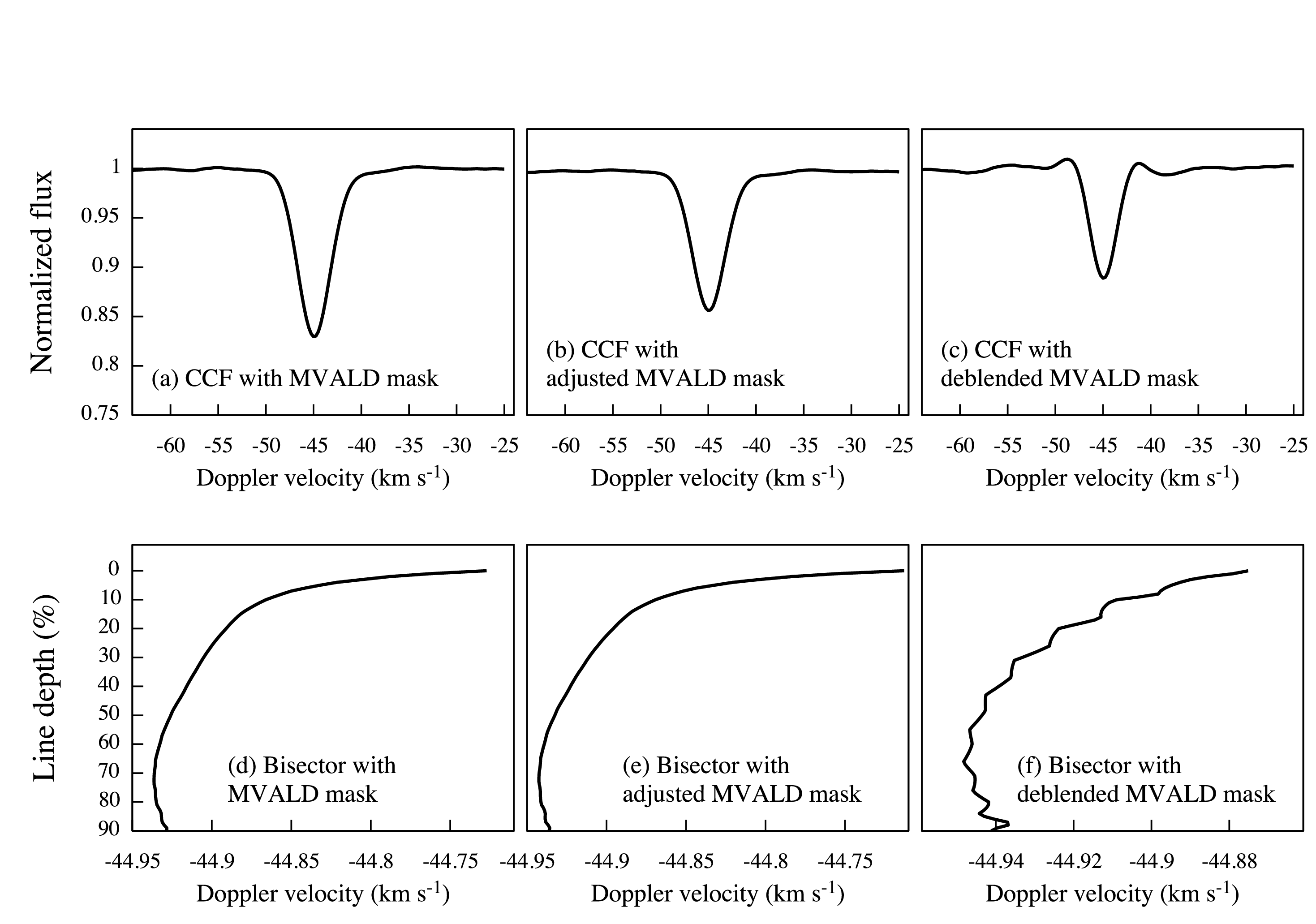}}
   \caption{Examples of CCFs (upper row) and bisectors (lower row) obtained with the MVALD (left),
   adjusted MVALD (middle), and deblended MVALD (right) mask respectively.}
   \label{ccf_bisVALD}
   \end{figure*}

   Both masks give similar results in term of CCF and bisector's shape (see Fig.~\ref{ccf_bisVALD}),
   and RV precision, with the MVALD mask performing slightly
   better (see Tab.~\ref{vrad_bis_rms}).

   As an additional test, we eliminated from the MVALD mask all the lines that
   fall in the same resolution step with each other (i.e., the lines where
   ${\lambda}_{2} - {\lambda}_{1} < \lambda/R$), creating an almost deblended
   mask. The results are definitely worse: the larger rms (see Tab.~\ref{vrad_bis_rms})
   may be due to the lower number of lines in the deblended mask (6501 against the
   original 17547), while 
   we found no explanation on the appearance of both the CCF's bumps and the
   bisector's segmentation (see Fig.~\ref{ccf_bisVALD}).

\begin{figure*}
   \subfloat{\includegraphics[width=0.5\textwidth]{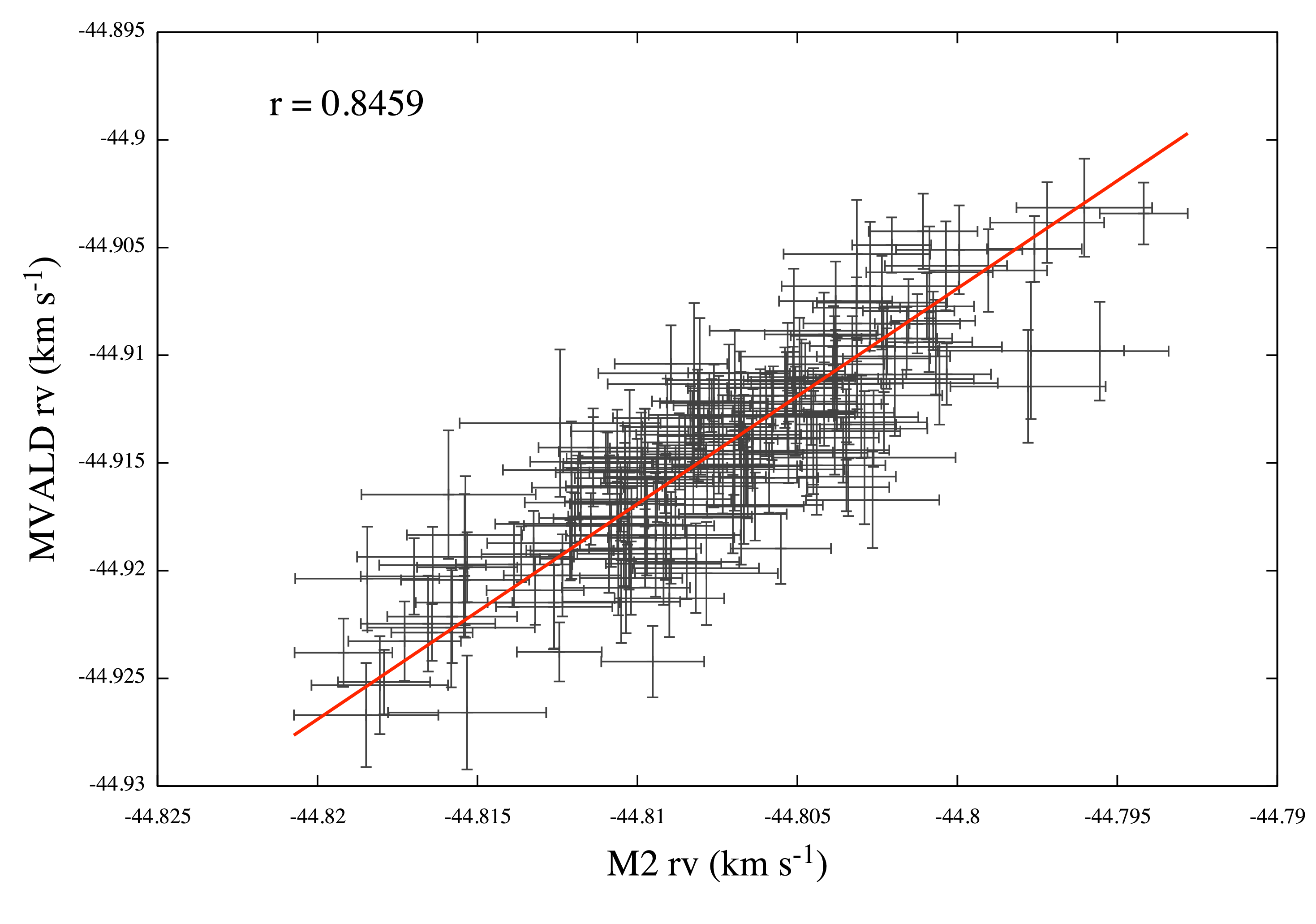}}
   \subfloat{\includegraphics[width=0.5\textwidth]{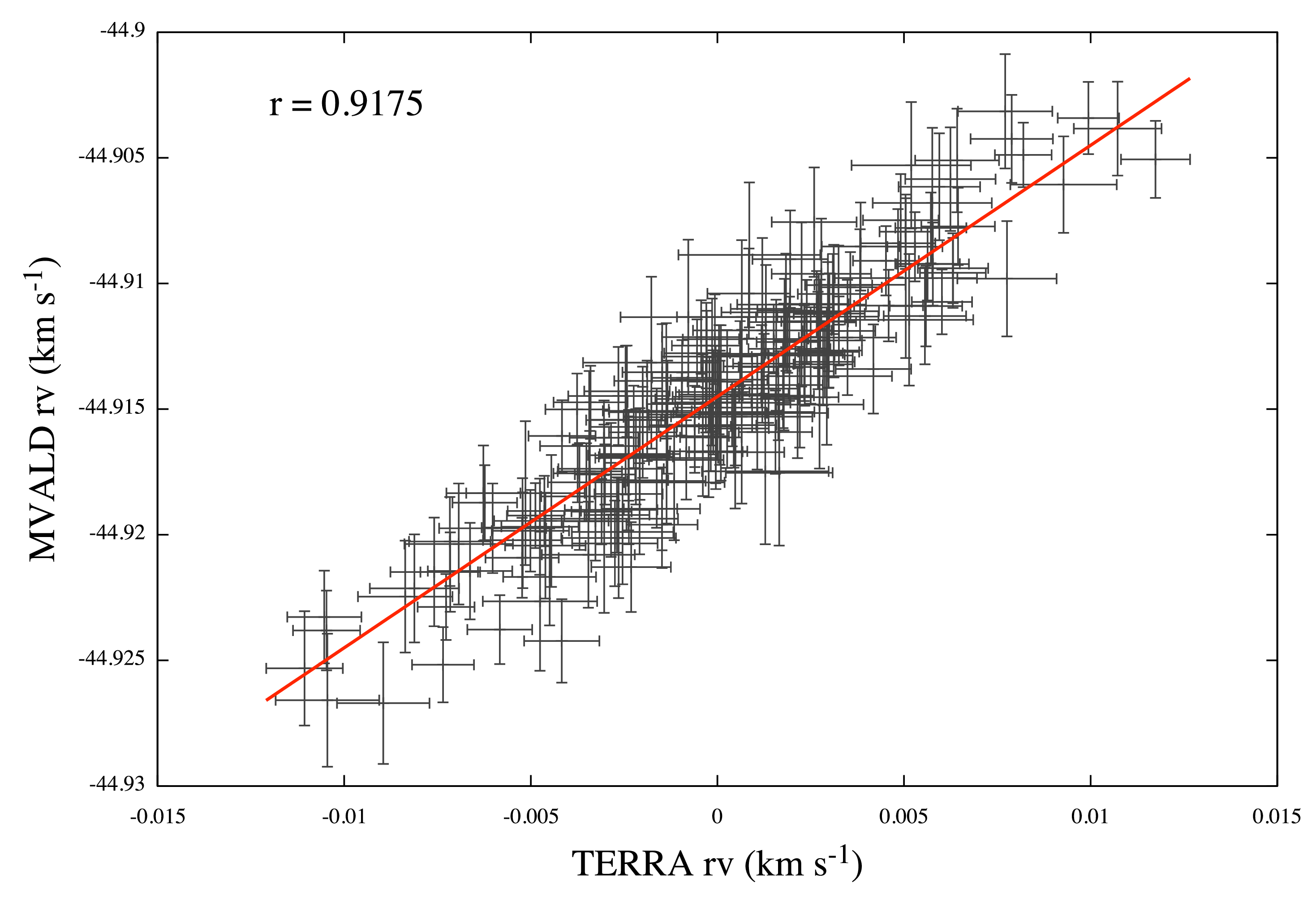}}
   \caption{\textit{Left:} M2 and MVALD RVs of GJ~3998. The red line show the
   one-to-one correlation passing through the median values of the RVs. The
   correlation coefficient is $r=0.8459$.
   \textit{Right:} HARPS-TERRA and MVALD RVs of GJ~3998. The RVs derived
   with HARPS-TERRA are centered around 0 km s$^{-1}$. The red line show the
   one-to-one correlation passing through the median values of the RVs. The
   correlation coefficient is $r=0.9175$.}
   \label{rv_corr}
   \end{figure*}

   We selected the MVALD mask as our new working mask for M-type stars,
   and compared the RVs computed by the DRS using the MVALD mask
   with the results from both the M2 mask and HARPS-TERRA: we found them to be in good agreement
   (see Fig.~\ref{rv_corr}), but the MVALD RVs correlate better with the HARPS-TERRA
   results.
   The RV errors found with HARPS-TERRA are however quite smaller (in the case of
   GJ~3998 the mean error from HARPS-TERRA is about 1.15 m s$^{-1}$, while the mean error using
   the MVALD mask is about 1.92 m s$^{-1}$).

\section{Search for exoplanets} \label{exosearch}
\label{sec:5}
   We tested the RV time-series obtained with the MVALD mask against those obtained
   with the M2 mask and with HARPS-TERRA by searching for already known exoplanets.

   The first step of the RV data analysis consists in identifying significant periodic signals in
   the data. Pre-whitening is a commonly used tool for finding multi-periodic signals in time-series
   data. With this method we find sequentially the dominant Fourier components in the time-series and
   remove them. The pre-whitening procedure was applied using the Generalized Lomb-Scargle (GLS)
   periodogram algorithm \citep{2009A&A...496..577Z} to the RV data obtained with the M2 and MVALD masks.
   We performed the same analysis also on the detrended RV data, obtained removing the correlation
   between RV and the bisector's span.
   We analyzed with the same method the time-series of bisector's span, 
   CCF contrast and FWHM, as well, to discriminate also signals possibly due to activity.

   We show the results of our work in Tab.~\ref{gls_table}: we analysed the exoplanet host stars
   GJ~3998 (confirmed planet with period $P=13.7d$), GJ~625 (confirmed planet with $P=14.6d$),
   GJ~686 (confirmed planet with $P=15.5d$), and GJ~3942 (confirmed planet with $P=6.9d$).

   All three methods find the signals due to the planetary orbits with compatible periods.

   \begin{table}
   \caption{Summary results of GLS analysis with different RV time-series}             % title of Table
   \label{gls_table}      % is used to refer this table in the text
   \centering                          % used for centering table
   \begin{tabular}{llll}        % centered columns (4 columns)
\hline\noalign{\smallskip}
   Mask & Period & Amplitude & FAP \\    % table heading
    & (days) & (m s$^{-1}$) & \\    % table heading 
\noalign{\smallskip}\hline\noalign{\smallskip}
    \multicolumn{3}{c}{GJ~3998, confirmed planet with $P=13.7d$} \\
\noalign{\smallskip}\hline
      M2 & 13.718 $\pm$ 0.009 & 2.68 $\pm$ 0.36 & $1.12\times 10^{-11}$\\      % inserting body of the table
      MVALD & 13.736 $\pm$ 0.008 & 2.8 $\pm$ 0.35 & $7.20\times 10^{-13}$ \\
      HARPS-TERRA & 13.729 $\pm$ 0.008 & 2.5 $\pm$ 0.3 & $2.39\times 10^{-12}$ \\
\noalign{\smallskip}\hline
   \multicolumn{3}{c}{GJ~625, confirmed planet with $P=14.6d$}\\
\noalign{\smallskip}\hline
      M2 & 14.627 $\pm$ 0.01 & 1.51 $\pm$ 0.25 & $2.24\times 10^{-7}$ \\      % inserting body of the table
      MVALD & 14.612 $\pm$ 0.009 & 2.2 $\pm$ 0.33 & $2.65\times 10^{-9}$ \\
      HARPS-TERRA & 14.612 $\pm$ 0.0075 & 1.8 $\pm$ 0.2 & $3.23\times 10^{-12}$ \\
\noalign{\smallskip}\hline
   \multicolumn{3}{c}{GJ~686, confirmed planet with $P=15.5d$}\\
\noalign{\smallskip}\hline
      M2 & 15.57 $\pm$ 0.02 & 2.8 $\pm$ 0.5 & $8.91\times 10^{-6}$ \\      % inserting body of the table
      MVALD & 15.501 $\pm$ 0.016 & 3.2 $\pm$ 0.5 & $1.13\times 10^{-6}$ \\
      HARPS-TERRA & 15.519 $\pm$ 0.014 & 2.8 $\pm$ 0.4 & $3.77\times 10^{-8}$ \\
\noalign{\smallskip}\hline
   \multicolumn{3}{c}{GJ~3942, confirmed planet with $P=6.9d$}\\
\noalign{\smallskip}\hline
      M2 & 6.906 $\pm$ 0.003 & 3.4 $\pm$ 0.6 & $1.61\times 10^{-7}$ \\      % inserting body of the table
      MVALD & 6.905 $\pm$ 0.003 & 3.1 $\pm$ 0.5 & $2.44\times 10^{-7}$ \\
      HARPS-TERRA & 6.905 $\pm$ 0.003 & 3.3 $\pm$ 0.5 & $3.99\times 10^{-10}$ \\
\noalign{\smallskip}\hline
   \end{tabular}
   \end{table}

If we investigate the significance of the signal, e.g. by comparing the false alarm
probabilities with the different methods (FAP, \cite{2009A&A...496..577Z}),
we can split our targets in two groups, taking into account whether activity signals
are found in the time-series or not:

\begin{itemize}
\item{GJ~625 and GJ~686: no activity signal is found in any of the time-series. The FAP of
the planetary signal with the MVALD mask falls between the results from HARPS-TERRA and the M2 mask;}
\item{GJ~3998 and GJ~3942: the activity signal is found in all the RV time-series.
In both cases the FAP of the activity signal is highest in the MVALD data, and the value
increases when detrending the RV time-series using the bisector's span time-series.
The situation regarding the planetary signal is less clear: while in the case of GJ~3998
the MVALD mask (both original and detrended) gives the lowest FAP value, in the case of GJ~3942
the MVALD mask performs worse.
}
\end{itemize}

There is still not enough statistics to confirm the usefulness of the bisector's span
as computed with the MVALD mask in the modeling and/or removing of the activity signal from the
RV time-series, but the MVALD mask do perform better than the standard M2 mask on recovering the
planetary signal in our sample of data.
The MVALD bisector's shape (see Fig.~\ref{ccf_bisVALD}$d$) is very similar to the K5 one
(see Fig.~\ref{ccfbis01}$e$) and it does not show any obvious anomaly.

%-----------------------------------------------------------------

\section{Conclusions}
\label{sec:6}
   HARPS and HARPS-N are still among the leading instruments in the RV exoplanet
   search, and their pipelines usually give very accurate results regarding 
   both the radial velocity and the activity indicators derived from the CCF.
   Unfortunately, this is not the case for M-type stars: the default M2 mask
   gives a strangely shaped CCF, and the K5 mask results in large errors bars
   on the extracted RVs.
   By manipulating the M2 mask we could obtain better results
   in the RV computation, but with some underlying problems
   still remaining: in particular, the shape of the CCF's bisector seems
   to have little physical meaning. 

   M-type stars are very interesting targets for exoplanet studies, but their
   high levels of activity have to be taken into account. Spectroscopic studies
   rely on several activity indicators, but the faintness of these objects
   favors the use of indicators derived from the CCF.
   The chief amongst those is the bisector's span, that is directly derived
   from the bisector of the CCF.
   Given that both the original and the modified M2 mask yielded bisectors
   with shapes devoid of an apparent physical meaning, we created new stellar masks
   using the line list provided by the VALD3 database. We called the new masks MVALD, adjusted MVALD
   (where the line depths were adjusted using an observed spectrum), and deblended MVALD
   (without the line blends).

   Working with the spectroscopic time-series of two early M-type exoplanet hosting stars
   (GJ~3998 and GJ~625), we found that the best results
   concerning both the RVs precision and the bisector's shape are
   obtained using the original MVALD mask, while the deblended MVALD mask
   shows the beginning of the segmentation behaviour of the bisector.
   We found good agreement between the RVs found with the MVALD mask
   and those from HARPS-TERRA, even if the latter have smaller errors. However HARPS-TERRA does
   not produce a CCF, which is the main limitation of the software, in particular when
   observing active stars. With the MVALD mask we recovered the signals of the known
   exoplanets orbiting around GJ~3998, GJ~625, GJ~686, and GJ~3942 with results comparable
   with those of HARPS-TERRA.
   Investigating the significance of the signals using the FAP values show that the
   results obtained using the MVALD mask tend to fall in-between those from HARPS-TERRA and
   the M2 mask.

   Now that the search for exoplanets is moving to targets different
   from solar-like stars it is important to use new masks for the CCF computation
   instead of relying on the standard ones: the MVALD mask we created may be a very
   useful tool for the radial velocity and activity index computation of early  M-type stars.

\begin{acknowledgements}
M.R. acknowledges financial support from INAF through the competitive project "FRONTIERA-2016".
The authors thank the following people for useful comments and insights: J. Maldonado Prado,
M. Pinamonti, A. Bignamini, G. Bruno, S. Benatti, E. Poretti.
\end{acknowledgements}


\bibliographystyle{spbasic}      % basic style, author-year citations
\bibliography{masksM.bib}   % name your BibTeX data base

%\begin{thebibliography}{}
% etc

%\bibitem[Affer et al.(2016)]{2016A&A...593A.117A} Affer, L., Micela, G., Damasso, M., et al.\ 2016, A\&A, 593, A117

%\end{thebibliography}

\end{document}